\documentclass[prl,twocolumn,amsmath,amssymb,superscriptaddress]{revtex4}

\usepackage{color}

\usepackage{graphicx}
\usepackage{graphics}
\usepackage{dcolumn}
\usepackage{bm}

\begin{document}

\title{Probing magnetism in the vortex phase of PuCoGa$_5$ by X-ray magnetic circular dichroism}

\author{N. Magnani}
\affiliation{European Commission, Joint Research Centre (JRC), Directorate for Nuclear Safety and Security, Postfach 2340, D-76125 Karlsruhe, Germany}

\author{R. Eloirdi}
\affiliation{European Commission, Joint Research Centre (JRC), Directorate for Nuclear Safety and Security, Postfach 2340, D-76125 Karlsruhe, Germany}

\author{F. Wilhelm}
\affiliation{European Synchrotron Radiation Facility (ESRF), B.P.220, F-38043 Grenoble, France}

\author{E. Colineau}
\affiliation{European Commission, Joint Research Centre (JRC), Directorate for Nuclear Safety and Security, Postfach 2340, D-76125 Karlsruhe, Germany}

\author{J.-C. Griveau}
\affiliation{European Commission, Joint Research Centre (JRC), Directorate for Nuclear Safety and Security, Postfach 2340, D-76125 Karlsruhe, Germany}

\author{A. B. Shick}
\affiliation{Institute of Physics, ASCR, Na Slovance 2, CZ-18221
Prague, Czech Republic}

\author{G. H. Lander}
\affiliation{European Commission, Joint Research Centre (JRC), Directorate for Nuclear Safety and Security, Postfach 2340, D-76125 Karlsruhe, Germany}

\author{A. Rogalev}
\affiliation{European Synchrotron Radiation Facility (ESRF), B.P.220, F-38043 Grenoble, France}

\author{R. Caciuffo}
\affiliation{European Commission, Joint Research Centre (JRC), Directorate for Nuclear Safety and Security, Postfach 2340, D-76125 Karlsruhe, Germany}

\date{September 14, 2017}

\begin{abstract}

We have measured X-ray magnetic circular dichroism (XMCD) spectra at the Pu $M_{4,5}$ absorption edges from a newly-prepared high-quality single crystal of the heavy fermion superconductor $^{242}$PuCoGa$_{5}$, exhibiting a critical temperature $T_{c} = 18.7~{\rm K}$. The experiment probes the vortex phase below $T_{c}$ and shows that an external magnetic field induces a Pu 5$f$ magnetic moment at 2 K equal to the temperature-independent moment measured in the normal phase up to 300 K by a SQUID device. This observation is in agreement with theoretical models claiming that the Pu atoms in PuCoGa$_{5}$ have a nonmagnetic singlet ground state resulting from the hybridization of the conduction electrons with the intermediate-valence 5$f$ electronic shell. Unexpectedly, XMCD spectra show that the orbital component of the $5f$ magnetic moment increases significantly between 30 and 2 K; the antiparallel spin component increases as well, leaving the total moment practically constant. We suggest that this indicates a low-temperature breakdown of the complete Kondo-like screening of the local 5$f$ moment.
\end{abstract}

\maketitle

PuCoGa$_{5}$ is a prototypical heavy-fermion compound that becomes a superconductor below  $T_{c} \simeq$ 18.5 K \cite{sarrao02}, the highest critical temperature of any heavy-fermion material. Fifteen years on its discovery, the nature of the pairing boson in PuCoGa$_{5}$ remains an open question. Superconductivity (SC) mediated by spin fluctuations (SFs) associated with the proximity to an antiferromagnetic (AFM) quantum critical point (QCP) was initially proposed. The hypothesis was supported by the observation in the normal phase of a Curie-Weiss (CW) behavior of the magnetic susceptibility $\chi_m$, suggesting the presence of Pu atoms carrying a local magnetic moment. Further arguments in favor of SFs-controlled SC were provided by NMR studies \cite{curro05}, revealing a nodal-gap function separating the condensate from the unpaired states. Subsequent point-contact spectroscopy measurements confirmed that the wavefunction of the paired electrons has an unconventional $d$-wave symmetry
\cite{daghero12}.
However, the SF conjecture was questioned \cite{jutier08,ummarino09} after polarized neutron diffraction failed to observe a local magnetic moment in the normal state of PuCoGa$_{5}$ \cite{hiess08}, pointing to an extrinsic origin of the reported temperature dependent $\chi_m$. This observation is confirmed in the present article by showing that the magnetic susceptibility of an almost defect-free PuCoGa$_{5}$ single crystal is weak and temperature-independent from $T_{c}$ up to room temperature.

Other members of the Pu$MX_{5}$ family ($M$ = Co, Rh, and $X$ = Ga, In) also become superconductors, with
$T_c$ ranging from $\sim$1.7 K in the case of PuRhIn$_{5}$ to $\sim$8.7 K for PuRhGa$_{5}$ \cite{bauer15}. The much larger $T_{c}$ of PuCoGa$_{5}$ could indicate that a different pairing mechanism is acting in the various compounds of the family. Indeed, Bauer \textit{et al.} \cite{bauer12} proposed that SC in PuCoIn$_{5}$ ($T_{c}$ = 2.5 K) is related to an AFM QCP, whereas  PuCoGa$_{5}$ would reside on a larger SC dome in the temperature-pressure (hybridization strength) phase diagram, around a valence fluctuation (VF) QCP. The hypothesis is supported by dynamical mean field theory (DMFT) calculations resulting in a quasiparticle peak at the Fermi level that is sharper in PuCoIn$_{5}$ than in PuCoGa$_{5}$, which suggests a more localized 5$f$-electron character in the less dense $X$ = In compound \cite{zhu12}. For $X$ = Ga, the
Pu atom would be in an intermediate-valence ground state between $5f^{5}$ and $5f^{6}$, with a fractional occupation number $n_{f} \sim 5.2$ \cite{pezzoli11}. Similar conclusions are reached by electronic structure calculations combining the density-functional theory (DFT) with an exact diagonalization (ED) of the Anderson impurity model \cite{shick13}. For $X$ = Ga, these calculations provide an electron density of states in reasonable agreement with photoemission measurements \cite{joyce03,eloirdi09}  and a non-magnetic singlet for the plutonium ground state. In this framework, the 5$f$ local magnetic moment is quenched by the combination of intermediate valence and hybridization with the surrounding cloud of conduction electrons. On the other hand, for $X$ = In, the predicted Pu ground state is magnetic as a result of a weaker hybridization strength \cite{shick13}.

The occurrence of valence fluctuations in PuCoGa$_{5}$ has been recently suggested by resonant ultrasound spectroscopy measurements, showing that the three compressional elastic moduli exhibit anomalous softening upon cooling, which is truncated at the SC transition \cite{ramshaw15}. These results have been interpreted as evidence for a valence transition at a $T_{V} < T_{c}$ that is avoided by the superconducting state
\cite{ramshaw15}. On the other hand, the relaxation rate isotope ratio T$_{1}^{-1}$($^{71}$Ga)/T$_{1}^{-1}$($^{69}$Ga) provided by
Nuclear Quadrupole Resonance
is not compatible with the presence of charge fluctuations in the normal state, but rather indicates the presence of anisotropic SFs \cite{koutroulakis16} that could, nevertheless, be associated with charge (valence) fluctuations with a higher energy scale.
High-resolution powder x-ray-diffraction recently showed that the volume expansion of PuCoGa$_5$ deviates from the curve expected for a simple Gr\"uneisen-Einstein model, but the observed variations are too small to be taken as an indication for the proximity of the system to a valence instability \cite{eloirdi17}.
The origin of SC in PuCoGa$_{5}$ remains therefore unclear. Alternative models have also been proposed, for instance assuming a \textit{composite} pairing in a lattice of Kondo ions screened by two distinct channels \cite{flint08,flint10}, or interband pairing with a sign-changing gap driven by SFs arising from spin-orbit split 5$f$ states and 5$f$-5$f$ and 3$d$-5$f$ particle-hole transitions \cite{graf15}.

To shed further light on the extraordinary properties of PuCoGa$_{5}$ we have measured X-ray magnetic circular dichroism (XMCD) spectra at the Pu $M_{4,5}$ absorption edges from a newly-prepared high-quality single crystal of this material.
The XMCD experiment was performed at the ID12 beamline \cite{rogalev01} of the ESRF
in Grenoble. Data have been collected between 2 and 30 K on a single crystal sample (with approximate size 2.5 $\times$ 1.0 $\times$ 0.05 mm) grown in a Ga flux at the Karlsruhe establishment of the JRC. The sample was prepared using
$^{242}$Pu metal obtained by amalgamation process
to avoid effects from radiation damage and self-heating.
The isotopic composition of the PuCoGa$_5$ sample used for the experiment (99.99 wt\% $^{242}$Pu, 0.0009 wt\% $^{241}$Pu, 0.0063 wt\% $^{240}$Pu, 0.0021 wt\% $^{239}$Pu, 0.00057 wt\% $^{238}$Pu on October 2014) was checked by ICP-MS.
The sample mass was 1.00~mg, corresponding to a plutonium mass of 0.37~mg and an activity of $\sim$54~kBq. The crystal was glued with Stycast$^{\circledR}$ 1266 transparent epoxy resin on an aluminum holder, with the crystallographic $c$-axis parallel to the incident X-ray beam and to the applied magnetic field.  The sample holder was then introduced into a hermetic Al capsule with two Kapton\texttrademark~windows of 62 $\mu$m thickness in total, following a protocol developed for XMCD measurements on other transuranium elements \cite{halevy12,magnani15}. In addition, magnetic susceptibility measurements were also carried out, in the temperature range 2$-$300 K, with an external magnetic field up to 7 T on a 697 mg sample using the MPMS-7 superconducting quantum interference device (SQUID) from Quantum Design available at JRC-Karlsruhe.

\begin{figure}
\centerline{ \includegraphics[width=7.5cm]{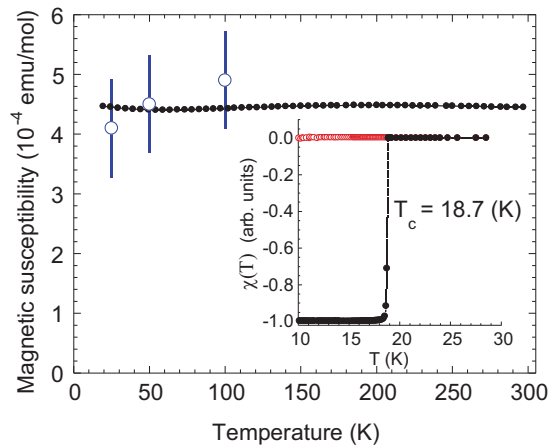}}
\caption{(Color online) Magnetic susceptibility in the normal state of PuCoGa$_{5}$ measured with an applied field of 1 mT on warming the sample after zero-field cooling (filled black dots). The blue open circles represent values deduced from polarized neutron diffraction measurements \cite{hiess08}. Inset: Temperature dependence of the magnetic susceptibility measured under zero-field cooling (filled black dots) and field cooling conditions (open red dots) in an applied field of 1 mT, providing $T_{c} = 18.7~{\rm K}$. \label{suscept}}
\end{figure}

The SQUID susceptibility curves for the investigated sample are shown in Fig. \ref{suscept}. From these data one obtains a critical temperature $T_{c}$ = 18.7 K (confirmed by heat capacity measurements not shown here). Contrary to magnetization measurements reported in earlier papers, but in agreement with neutron scattering results \cite{hiess08}, the magnetic susceptibility in the normal phase, $\chi_{m}$, is practically temperature independent between $T_{c}$ and room temperature. This is the typical behavior of intermediate-valence systems well below the characteristic charge fluctuation temperature $T_{fc}$ \cite{khomskii79}.

X-ray absorption spectroscopy (XAS) and XMCD data have been collected at several temperatures in the photon energy range between 3720 and 4040 eV, across the $M_{4,5}$ edges of Pu.
The XAS spectra were recorded in backscattering geometry using the total-fluorescence-yield detection mode. The beam intensity was measured for parallel $\mu^{+}(E)$ and antiparallel $\mu^{-}(E)$ photon helicity, in a magnetic field $B_{\parallel} = 17~{\rm T}$. The XAS, ($\mu^{+}(E)+\mu^{-}(E))/2$, and the XMCD spectra, $\mu^{+}(E)-\mu^{-}(E)$, were obtained after applying self-absorption and incomplete polarization corrections using standard procedures discussed in \cite{wilhelm13}. Any variation of the irradiated volume (for example due to sample motion) is
corrected by normalizing the spectra to the edge jump. In the superconducting phase the magnetic field penetrates into the sample forming vortices. The XMCD signal is different from zero only if the atomic shells are polarized by the applied field, therefore only the vortex cores of the superconducting state contribute to it. On the other hand, the XANES signal is not affected by the superconducting transition. This means that  XMCD provides atomic quantities averaged over all plutonium atoms in the irradiated volume both above and below $T_c$.

The penetration depth for PuCoGa$_{5}$ at 2 K and $B = 60~{\rm mT}$ is $\lambda  = 265~{\rm nm}$ \cite{ohishi07} and is bound to increase for larger fields \cite{sonier97}, whereas the Ginzburg-Landau coherence length is $\xi \sim 2.1~{\rm nm}$ \cite{sarrao02}. These values must be compared with the penetration of the X-ray beam at the $M_{4}$ edge, which is $\sim 200$ nm. Assuming a critical field $B_{c2} = 63(1-T^{2}/T_{c}^{2})$ \cite{ummarino09b} and $\kappa = \lambda/\xi = 126$, the temperature variation of the volume average of the magnetic field
is less than 0.1\%.  We can therefore be confident that any temperature dependence of quantities probed by XMCD is not related to changes in the flux line lattice.

\begin{figure}
\includegraphics[width=8.0cm]{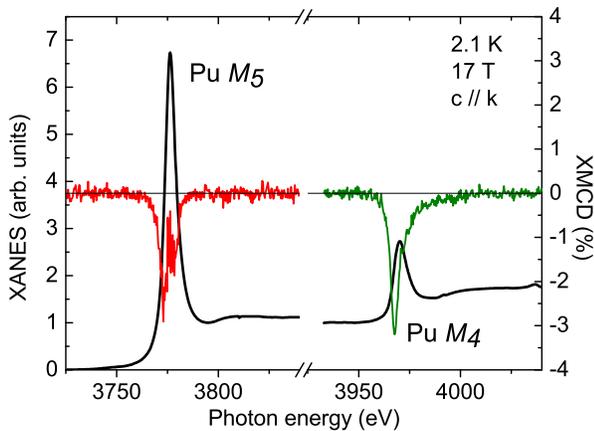}
\caption{(Color online) The X-ray absorption near-edge structure (XANES, solid black lines) and X-ray magnetic circular dichroism (XMCD) spectra as a function of photon energy through the Pu $M_5$ (red line) and $M_4$ (green line) edges in PuCoGa$_5$. \label{XASXMCD}}
\end{figure}

After cooling the sample to 2.1 K in zero magnetic field, $B_{\parallel}$ was applied and data were collected at several
temperatures up to 30 K (according to \cite{ummarino09b}, $T_{c} \sim 15.4~{\rm K}$ for $B_{\parallel} = 17~{\rm T}$).
The spectra at 2.1 K are shown in Fig.~\ref{XASXMCD}.
The XAS branching ratio $B = I_{M_{5}}/(I_{M_{5}}+I_{M_{4}})$ is proportional to the expectation value of the angular part of the valence states spin-orbit operator $2 \langle {\bf l} \cdot {\bf s} \rangle = 3 n_{7/2} - 4 n_{5/2}$ \cite{thole88},
\begin{equation}\label{so}
\frac{2\langle {\bf l} \cdot {\bf s} \rangle}{3n_{h}}-\Delta = -\frac{5}{2}(B-\frac{3}{5})
\end{equation}
where $n_h = 14 - n_f$ is the number of holes in the 5$f$ shell, $I_{M_{4,5}}$ is the integrated intensity of the isotropic X-ray absorption spectra at the $M_{4,5}$ edge, and
$\Delta$ is a quantity dependent from the electronic configuration,
which we will neglect here since it is equal to zero for Pu$^{3+}$
\cite{vanderlaan04}. No appreciable temperature variation is observed for the branching ratio. Inserting in Eq. \ref{so} the experimental values at 2 K, $I_{M_{5}}$ = 51.04(8) and $I_{M_{5}}+I_{M_{4}}$ = 63.5(1), we find $B = 0.804(3)$, which within experimental errors coincides with the value measured by XAS for PuFe$_{2}$ \cite{wilhelm13} and by electron energy-loss spectroscopy for $\alpha$-plutonium \cite{vanderlaan04}. It is also close to the value expected for a $5f^5$ configuration assuming intermediate coupling (IC) ($B = 0.83 $) \cite{vanderlaan04} and slightly smaller than the value measured for PuSb ($B = 0.848(8)$) \cite{janoschek15}.

The orbital contribution to the magnetic moment carried by the Pu atoms can be determined as \cite{thole92}
\begin{equation}\label{om}
\langle L_{z}\rangle = \frac{n_h}{I_{M_{5}}+I_{M_{4}}} (\Delta I_{M_{5}}+\Delta I_{M_{4}})
\end{equation}
where $\Delta I_{M_{4,5}}$ is the partial integrated dichroic signal at the Pu $M_{4,5}$ edge. Applying this sum rule to the spectra recorded at 30 K in a 17-tesla field ($\Delta I_{M_{5}}$ = -0.16(1); $\Delta I_{M_{4}}$ = -0.21(1)), we obtain the orbital moment on Pu as $\mu_L=-\langle L_{z}\rangle$ = + 0.052(2) $\mu_B$. Interestingly, from the spectra measured at 2 K ($\Delta I_{M_{5}}$ = -0.20(1); $\Delta I_{M_{4}}$ = -0.28(1)) we obtain a slightly larger induced orbital moment of + 0.068(2) $\mu_B$.

\begin{table*}
\caption{Values of the moments and moment ratios obtained from the analysis of our XMCD data compared with those calculated by DFT+ED. The quantities which can be directly determined by the XMCD experiment without any assumptions other than the validity of the sum rules are $\langle L_z \rangle$ and $\langle S_{\rm eff} \rangle$, which are given in the first two lines. The error bars given in parentheses refer to the least significant digit. \label{summary}}
\begin{center}
\begin{tabular}{cddddd}
\hline
\hline
Quantity & \multicolumn{2}{c}{Assuming $r = r_{\rm IC}$} & \multicolumn{2}{c}{Assuming $\mu = \mu_{\rm SQUID}$} & \multicolumn{1}{c}{Calculated} \\
(units) & \multicolumn{1}{r}{$T$ = 2 K} & \multicolumn{1}{r}{$T$ = 30 K} & \multicolumn{1}{r}{$T$ = 2 K} & \multicolumn{1}{r}{$T$ = 30 K} & \multicolumn{1}{c}{(DFT+ED)} \\
\hline
$\mu_L = -\langle L_z \rangle$ ($\mu_B$) & +0.068(2) & +0.052(2) & +0.068(2) & +0.052(2) & +0.048 \\
$\langle S_{\rm eff} \rangle = \langle S_z \rangle + 3 \langle T_z \rangle$ ($\mu_B$) & +0.016(1) & +0.011(1) & +0.016(1) & +0.011(1) & +0.010 \\
$\mu_S = -2\langle S_z \rangle$ ($\mu_B$) & -0.040(3) & -0.028(3) & -0.053(8) & -0.037(2) & -0.067 \\
$\mu = \mu_L + \mu_S$ ($\mu_B$) & +0.028(4) & +0.024(4) & +0.015(8) & +0.015(1) & -0.019 \\
$R = \mu_L / \mu_S$ & -1.7(1) & -1.9(2) & -1.3(2) & -1.4(1) & -0.72 \\
$r = 3 \langle T_z \rangle / \langle S_z \rangle$ & -0.218 & -0.218 & -0.41(8) & -0.41(8) & -0.72 \\
\hline
\hline
\end{tabular}
\end{center}
\end{table*}

A second sum rule correlates the measured dichroic signal and the spin polarization $\langle S_{z} \rangle$, stating that
\cite{carra93}
\begin{equation}\label{sm}
\langle S_{\rm eff} \rangle \equiv \langle S_{z}\rangle + 3\langle T_{z}\rangle = \frac{n_{h}}{2(I_{M_{5}}+I_{M_{4}})} (\Delta I_{M_{5}}-\frac{3}{2}\Delta I_{M_{4}}).
\end{equation}
Therefore, in order to determine the spin component of the magnetic moment ($\mu_S = -2 \langle S_z \rangle$) from XMCD measurements it is necessary to extract the value of $\langle T_z \rangle$, the $z$ component of the expectation value of the magnetic dipole operator $\bf{T} = \sum_i[\bf{s}_i-3\bf{r}_i(\bf{r}_i \cdot \bf{s}_i)/r_i^2]$. This can be done either by using a theoretical estimate for $\langle T_z \rangle$ or by combining XMCD with another experimental technique which provides the value of the total magnetic moment $\mu = \mu_L+\mu_S$. Each of these approaches has its advantages and disadvantages, some of which will be clarified below. However, an inspection of Table \ref{summary} shows that in this case one obtains the same qualitative result with both methods: the total $5f$ magnetic moment is temperature-independent even below $T_c$.

For the first approach, we assume that the ratio $r = 3 \langle T_z \rangle / \langle S_z \rangle$ has the temperature-independent value calculated in IC for a $5f^5$ configuration, $r_{\rm IC} = -0.218$, and use it to calculate $\langle T_z \rangle$ from the experimentally measured $\langle S_{\rm eff} \rangle$. At first one might argue that this simple choice is not suitable to describe the multiconfigurational ground state of PuCoGa$_5$, but considering that the weight of the $5f^4$ wavefunction is expected to be small and that the $5f^6$ states only contribute with very weak induced moments it actually appears to be a good approximation (a similar situation is found for example in the well-known intermediate-valence compound CePd$_3$, whose XMCD signal at the Ce-$L_{2,3}$ edges arises only from the $4f^1$ final state \cite{kappler04}). Nevertheless, analyzing our XMCD data with this method we obtain a value $\mu = 0.024(4)~\mu_B$
at 30 K, which agrees only qualitatively with the induced magnetic moment obtained by SQUID measurements at the same temperature, $\mu_{\rm SQUID} = \chi_m(T) \times B_{\parallel}$ (Fig. \ref{suscept}); in fact, taking into account the diamagnetic contributions for the argon (on the Co and Ga sites) and radon (on
the Pu site) core electrons, $\chi_d \approx 0.5 \times 10^{-4}$ emu/mol, we estimate that at 30 K $\mu_{\rm SQUID} =  0.015(1) ~ \mu_B$. Whereas in principle $\mu$ and $\mu_{\rm SQUID}$ cannot be compared directly, since the latter represents the total magnetic moment whereas the former only accounts for the $5f$-electron contribution, it must be remarked that the magnetic susceptibility at the Pu sites determined by neutron scattering is very close to the SQUID result. On the other hand, the most interesting result is that treating the data at 2 K leads to $\mu = 0.028(4)~\mu_B$, which coincides to the value determined at 30 K within the experimental uncertainties.

For the second approach, we assume that the value of $\mu$ measured by XMCD at 30 K equals $\mu_{\rm SQUID}$;
since $\mu_L = 0.052(2)~\mu_B$ is known from XMCD, we obtain $\mu_S = -0.037(2)~\mu_B$ and $r = -0.41(8)$. Although it does not exactly coincide with $r_{\rm IC}$, the value is quite reasonable for an electronic configuration close to 5$f^{5}$ \cite{magnani15}.
SQUID measurements cannot determine the magnetic moment below the SC transition temperature; however, $r$ is not expected to change with temperature so we can use its value
at 30 K to extract the Pu spin moment at 2 K from the XMCD measurements, obtaining $\mu_S = -0.053(8)~\mu_B$. The estimated total moment at 2 K is therefore $\mu = 0.015(8)~\mu_B$; despite the larger uncertainty, once again it turns out to be equal to that measured at 30 K.

The moments experimentally obtained with both approaches are listed in Table \ref{summary}. As explained above, $\mu_S$ and $\mu_L$ are averages over all irradiated plutonium atoms, and therefore their values represent lower limits for those inside the vortex phase. The Table also shows a comparison with selected results of DFT+ED calculations;
although some discrepancies between theory and experiment remain (for example, the calculations wrongly predict that the spin component would be larger than the orbital moment), the DFT+ED magnetic moments are in qualitative agreement with XMCD data, as opposed to the much larger values ($\mu_S=4.08~\mu_B$ and $\mu_L=-2.32~\mu_B$) calculated with DFT \cite{opahle03}.
It is also worth noticing that the ratio $R = \mu_L / \mu_S$ obtained with the $r = r_{\rm IC}$ assumption
is in very good agreement with the theory of Pezzoli {\it et al.} \cite{pezzoli11}. On the other hand, the large $\mu_L / \mu$ ratio obtained assuming $\mu = \mu_{\rm SQUID}$ would cause a ``hump'' in the neutron form factor which was not seen in the experiments \cite{hiess08}, although these neutron measurements were performed with the magnetic field applied in a different direction with respect to XMCD.

\begin{figure}
\includegraphics[width=8.0cm]{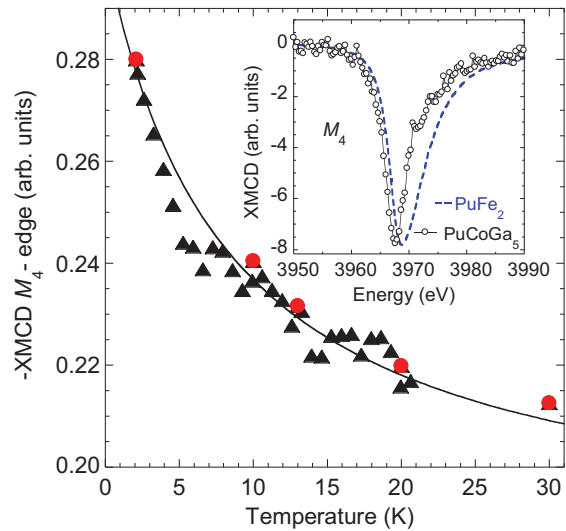}
\caption{(Color online) Temperature dependence of the XMCD signal measured at the $M_{4}$ absorption edge of Pu for PuCoGa$_5$, with a 17-T magnetic field applied along $c$. The red circles correspond to the integral of the XMCD spectra over the whole $M_4$-edge energy range, whereas the black triangles are the XMCD value measured at the $M_4$ peak energy (3968 eV). The line is a guide to the eye. Inset: XMCD spectra at the $M_{4}$ absorption edge of Pu measured for PuCoGa$_5$ (present work) and PuFe$_{2}$ \cite{wilhelm13}. The spectra are normalized to their respective maxima. Note that the signal measured on PuFe$_{2}$ was about 30 times larger. \label{MS}}
\end{figure}

Figure \ref{MS} shows the experimentally measured temperature dependence of the XMCD signal at the Pu $M_4$ edge. According to the sum rules $-\Delta I_{M_4}$ is proportional to $\mu_L - \mu_S$. This quantity increases monotonically from 30 to 2 K;
we therefore expect that the low-temperature increase observed for $\mu_L$ and $\left| \mu_S \right|$ (Table \ref{summary}) is monotonous as well. In particular, we note that
no clear anomaly is visible around $T_c$. Since the transition is second order, and therefore the volume fraction occupied by the vortex phase has no discontinuity, this means that also the magnetic moments change continuously.
The apparent discrepancy with the local spin susceptibility measured by NMR for PuCoGa$_5$ \cite{curro05}, which \emph{decreases} upon lowering the temperature below $T_c$, is due to the fact that these measurements probed the Co and Ga centers, whereas XMCD is sensitive only to the moments of the 5$f$ shell.
On the other hand, the increase of the Pu spin moment suggests that the Kondo-like screening required to reproduce the flat magnetic susceptibility in the normal phase partially breaks down at low temperature, possibly because of a change in the 5$f$ and conduction electron hybridization. De Luca {\it et al.} \cite{deluca10} have observed a similar effect in the spin susceptibility of high-$T_c$ superconductors, which they attributed to a field-induced reorientation of the fluctuating spins perpendicular to the CuO$_2$ planes.

The overall shape of the XMCD signal is also interesting. One single peak is observed at the $M_{4}$ edge, whith a full width at half maximum (FWHM) of about 5 eV. A narrow peak of similar width was also seen in PuSb \cite{janoschek15} but not in PuFe$_{2}$ where the $M_{4}$ peak is significantly broader (FWHM $\simeq$ 7.5 eV) and shifted by about 1 eV towards higher energy (Fig. \ref{MS}). The $M_{5}$ spectra show two peaks, separated by $\simeq$ 5 eV.
Again, the $M_{5}$ spectral shape for PuFe$_{2}$ is
different, being characterized by a sharp negative peak followed by a small positive upturn.
The close similarity between the spectral shape of PuCoGa$_{5}$ and PuSb (a well-known localized system), as well as the difference with PuFe$_{2}$ (a well-known itinerant system), means that we are probing the localized 5$f$ electron states in PuCoGa$_{5}$; as expected, treating its electronic states
as completely itinerant is an incorrect approximation.

We thank P. Colomp (ESRF radioprotection services) for his cooperation during the execution of the experiment, P. Amador-Celdran (JRC) for technical support, C. Brossard and M. Schulz (JRC) for their assistance in organizing the sample transport, and E. Zuleger (JRC) for the isotope analysis. A. B. S. acknowledges financial support provided by the Czech Science Foundation (GACR) grant No.~~15-07172S.

\bibliography{PuCoGa5}

\begin{thebibliography}{46}
\expandafter\ifx\csname natexlab\endcsname\relax\def\natexlab#1{#1}\fi
\expandafter\ifx\csname bibnamefont\endcsname\relax
  \def\bibnamefont#1{#1}\fi
\expandafter\ifx\csname bibfnamefont\endcsname\relax
  \def\bibfnamefont#1{#1}\fi
\expandafter\ifx\csname citenamefont\endcsname\relax
  \def\citenamefont#1{#1}\fi
\expandafter\ifx\csname url\endcsname\relax
  \def\url#1{\texttt{#1}}\fi
\expandafter\ifx\csname urlprefix\endcsname\relax\def\urlprefix{URL }\fi
\providecommand{\bibinfo}[2]{#2}
\providecommand{\eprint}[2][]{\url{#2}}

\bibitem[{\citenamefont{Sarrao et~al.}(2002)\citenamefont{Sarrao, Morales,
  Thompson, Scott, Stewart, Wastin, Rebizant, Boulet, Colineau, and
  Lander}}]{sarrao02}
\bibinfo{author}{\bibfnamefont{J.~L.} \bibnamefont{Sarrao}},
  \bibinfo{author}{\bibfnamefont{L.~A.} \bibnamefont{Morales}},
  \bibinfo{author}{\bibfnamefont{J.~D.} \bibnamefont{Thompson}},
  \bibinfo{author}{\bibfnamefont{B.~L.} \bibnamefont{Scott}},
  \bibinfo{author}{\bibfnamefont{G.~R.} \bibnamefont{Stewart}},
  \bibinfo{author}{\bibfnamefont{F.}~\bibnamefont{Wastin}},
  \bibinfo{author}{\bibfnamefont{J.}~\bibnamefont{Rebizant}},
  \bibinfo{author}{\bibfnamefont{P.}~\bibnamefont{Boulet}},
  \bibinfo{author}{\bibfnamefont{E.}~\bibnamefont{Colineau}}, \bibnamefont{and}
  \bibinfo{author}{\bibfnamefont{G.~H.} \bibnamefont{Lander}},
  \bibinfo{journal}{Nature (London)} \textbf{\bibinfo{volume}{420}},
  \bibinfo{pages}{297} (\bibinfo{year}{2002}).

\bibitem[{\citenamefont{Curro et~al.}(2005)\citenamefont{Curro, Caldwell,
  Bauer, Morales, Graf, Bang, Balatsky, Thompson, and Sarrao}}]{curro05}
\bibinfo{author}{\bibfnamefont{N.~J.} \bibnamefont{Curro}},
  \bibinfo{author}{\bibfnamefont{T.}~\bibnamefont{Caldwell}},
  \bibinfo{author}{\bibfnamefont{E.~D.} \bibnamefont{Bauer}},
  \bibinfo{author}{\bibfnamefont{L.~A.} \bibnamefont{Morales}},
  \bibinfo{author}{\bibfnamefont{M.~J.} \bibnamefont{Graf}},
  \bibinfo{author}{\bibfnamefont{Y.}~\bibnamefont{Bang}},
  \bibinfo{author}{\bibfnamefont{A.~V.} \bibnamefont{Balatsky}},
  \bibinfo{author}{\bibfnamefont{J.~D.} \bibnamefont{Thompson}},
  \bibnamefont{and} \bibinfo{author}{\bibfnamefont{J.~L.}
  \bibnamefont{Sarrao}}, \bibinfo{journal}{Nature (London)}
  \textbf{\bibinfo{volume}{434}}, \bibinfo{pages}{622} (\bibinfo{year}{2005}).

\bibitem[{\citenamefont{Daghero et~al.}(2012)\citenamefont{Daghero, Tortello,
  Ummarino, Griveau, Colineau, Eloirdi, Shick, Kolorenc, Lichtenstein, and
  Caciuffo}}]{daghero12}
\bibinfo{author}{\bibfnamefont{D.}~\bibnamefont{Daghero}},
  \bibinfo{author}{\bibfnamefont{M.}~\bibnamefont{Tortello}},
  \bibinfo{author}{\bibfnamefont{G.~A.} \bibnamefont{Ummarino}},
  \bibinfo{author}{\bibfnamefont{J.-C.} \bibnamefont{Griveau}},
  \bibinfo{author}{\bibfnamefont{E.}~\bibnamefont{Colineau}},
  \bibinfo{author}{\bibfnamefont{R.}~\bibnamefont{Eloirdi}},
  \bibinfo{author}{\bibfnamefont{A.~B.} \bibnamefont{Shick}},
  \bibinfo{author}{\bibfnamefont{J.}~\bibnamefont{Koloren\v{c}}},
  \bibinfo{author}{\bibfnamefont{A.~I.} \bibnamefont{Lichtenstein}},
  \bibnamefont{and} \bibinfo{author}{\bibfnamefont{R.}~\bibnamefont{Caciuffo}},
  \bibinfo{journal}{Nat. Commun.} \textbf{\bibinfo{volume}{3}},
  \bibinfo{pages}{786} (\bibinfo{year}{2012}).

\bibitem[{\citenamefont{Jutier et~al.}(2008)\citenamefont{Jutier, Ummarino,
  Griveau, Wastin, Colineau, Rebizant, Magnani, and Caciuffo}}]{jutier08}
\bibinfo{author}{\bibfnamefont{F.}~\bibnamefont{Jutier}},
  \bibinfo{author}{\bibfnamefont{G.~A.} \bibnamefont{Ummarino}},
  \bibinfo{author}{\bibfnamefont{J.-C.} \bibnamefont{Griveau}},
  \bibinfo{author}{\bibfnamefont{F.}~\bibnamefont{Wastin}},
  \bibinfo{author}{\bibfnamefont{E.}~\bibnamefont{Colineau}},
  \bibinfo{author}{\bibfnamefont{J.}~\bibnamefont{Rebizant}},
  \bibinfo{author}{\bibfnamefont{N.}~\bibnamefont{Magnani}}, \bibnamefont{and}
  \bibinfo{author}{\bibfnamefont{R.}~\bibnamefont{Caciuffo}},
  \bibinfo{journal}{Phys. Rev. B} \textbf{\bibinfo{volume}{77}},
  \bibinfo{pages}{024521} (\bibinfo{year}{2008}).

\bibitem[{\citenamefont{Ummarino
  et~al.}(2009{\natexlab{a}})\citenamefont{Ummarino, Magnani, Griveau,
  Rebizant, and Caciuffo}}]{ummarino09}
\bibinfo{author}{\bibfnamefont{G.}~\bibnamefont{Ummarino}},
  \bibinfo{author}{\bibfnamefont{N.}~\bibnamefont{Magnani}},
  \bibinfo{author}{\bibfnamefont{J.-C.} \bibnamefont{Griveau}},
  \bibinfo{author}{\bibfnamefont{J.}~\bibnamefont{Rebizant}}, \bibnamefont{and}
  \bibinfo{author}{\bibfnamefont{R.}~\bibnamefont{Caciuffo}},
  \bibinfo{journal}{J. Nucl. Mater.} \textbf{\bibinfo{volume}{385}},
  \bibinfo{pages}{4 } (\bibinfo{year}{2009}{\natexlab{a}}).

\bibitem[{\citenamefont{Hiess et~al.}(2008)\citenamefont{Hiess, Stunault,
  Colineau, Rebizant, Wastin, Caciuffo, and Lander}}]{hiess08}
\bibinfo{author}{\bibfnamefont{A.}~\bibnamefont{Hiess}},
  \bibinfo{author}{\bibfnamefont{A.}~\bibnamefont{Stunault}},
  \bibinfo{author}{\bibfnamefont{E.}~\bibnamefont{Colineau}},
  \bibinfo{author}{\bibfnamefont{J.}~\bibnamefont{Rebizant}},
  \bibinfo{author}{\bibfnamefont{F.}~\bibnamefont{Wastin}},
  \bibinfo{author}{\bibfnamefont{R.}~\bibnamefont{Caciuffo}}, \bibnamefont{and}
  \bibinfo{author}{\bibfnamefont{G.~H.} \bibnamefont{Lander}},
  \bibinfo{journal}{Phys. Rev. Lett.} \textbf{\bibinfo{volume}{100}},
  \bibinfo{pages}{076403} (\bibinfo{year}{2008}).

\bibitem[{\citenamefont{Bauer and Thompson}(1988)}]{bauer15}
\bibinfo{author}{\bibfnamefont{E.~D.} \bibnamefont{Bauer}} \bibnamefont{and}
  \bibinfo{author}{\bibfnamefont{J.~D.}~\bibnamefont{Thompson}},
  \bibinfo{journal}{Annu. Rev. Condens. Matter Phys.} \textbf{\bibinfo{volume}{6}},
  \bibinfo{pages}{137} (\bibinfo{year}{2015}).

\bibitem[{\citenamefont{Bauer et~al.}(2012)\citenamefont{Bauer, Altarawneh,
  Tobash, Gofryk, Ayala-Valenzuela, Mitchell, McDonald, Mielke, Ronning,
  Griveau et~al.}}]{bauer12}
\bibinfo{author}{\bibfnamefont{E.~D.} \bibnamefont{Bauer}},
  \bibinfo{author}{\bibfnamefont{M.~M.} \bibnamefont{Altarawneh}},
  \bibinfo{author}{\bibfnamefont{P.~H.} \bibnamefont{Tobash}},
  \bibinfo{author}{\bibfnamefont{K.}~\bibnamefont{Gofryk}},
  \bibinfo{author}{\bibfnamefont{O.~E.} \bibnamefont{Ayala-Valenzuela}},
  \bibinfo{author}{\bibfnamefont{J.~N.} \bibnamefont{Mitchell}},
  \bibinfo{author}{\bibfnamefont{R.~D.} \bibnamefont{McDonald}},
  \bibinfo{author}{\bibfnamefont{C.~H.} \bibnamefont{Mielke}},
  \bibinfo{author}{\bibfnamefont{F.}~\bibnamefont{Ronning}},
  \bibinfo{author}{\bibfnamefont{J.-C.} \bibnamefont{Griveau}},
  \bibinfo{author}{\bibfnamefont{E.} \bibnamefont{Colineau}},
  \bibinfo{author}{\bibfnamefont{R.} \bibnamefont{Eloirdi}},
  \bibinfo{author}{\bibfnamefont{R.}~\bibnamefont{Caciuffo}},
  \bibinfo{author}{\bibfnamefont{B.~L.} \bibnamefont{Scott}},
  \bibinfo{author}{\bibfnamefont{O.} \bibnamefont{Janka}},
  \bibinfo{author}{\bibfnamefont{S.~M.} \bibnamefont{Kauzlarich}}, \bibnamefont{and}
  \bibinfo{author}{\bibfnamefont{J.~D.} \bibnamefont{Thompson}},
  \bibinfo{journal}{J. Phys. Condens. Matter}
  \textbf{\bibinfo{volume}{24}}, \bibinfo{pages}{052206}
  (\bibinfo{year}{2012}).

\bibitem[{\citenamefont{Zhu et~al.}(2012)\citenamefont{Zhu, Tobash, Bauer,
  Ronning, Scott, Haule, Kotliar, Albers, and Wills}}]{zhu12}
\bibinfo{author}{\bibfnamefont{J.-X.} \bibnamefont{Zhu}},
  \bibinfo{author}{\bibfnamefont{P.~H.} \bibnamefont{Tobash}},
  \bibinfo{author}{\bibfnamefont{E.~D.} \bibnamefont{Bauer}},
  \bibinfo{author}{\bibfnamefont{F.}~\bibnamefont{Ronning}},
  \bibinfo{author}{\bibfnamefont{B.~L.} \bibnamefont{Scott}},
  \bibinfo{author}{\bibfnamefont{K.}~\bibnamefont{Haule}},
  \bibinfo{author}{\bibfnamefont{G.}~\bibnamefont{Kotliar}},
  \bibinfo{author}{\bibfnamefont{R.~C.} \bibnamefont{Albers}},
  \bibnamefont{and} \bibinfo{author}{\bibfnamefont{J.~M.} \bibnamefont{Wills}},
  \bibinfo{journal}{Europhys. Lett.)} \textbf{\bibinfo{volume}{97}},
  \bibinfo{pages}{57001} (\bibinfo{year}{2012}).

\bibitem[{\citenamefont{Pezzoli et~al.}(2011)\citenamefont{Pezzoli, Haule, and
  Kotliar}}]{pezzoli11}
\bibinfo{author}{\bibfnamefont{M.~E.} \bibnamefont{Pezzoli}},
  \bibinfo{author}{\bibfnamefont{K.}~\bibnamefont{Haule}}, \bibnamefont{and}
  \bibinfo{author}{\bibfnamefont{G.}~\bibnamefont{Kotliar}},
  \bibinfo{journal}{Phys. Rev. Lett.} \textbf{\bibinfo{volume}{106}},
  \bibinfo{pages}{016403} (\bibinfo{year}{2011}).

\bibitem[{\citenamefont{Shick et~al.}(2013)\citenamefont{Shick, Kolorenc, Rusz,
  Oppeneer, Lichtenstein, Katsnelson, and Caciuffo}}]{shick13}
\bibinfo{author}{\bibfnamefont{A.~B.} \bibnamefont{Shick}},
  \bibinfo{author}{\bibfnamefont{J.}~\bibnamefont{Koloren\v{c}}},
  \bibinfo{author}{\bibfnamefont{J.}~\bibnamefont{Rusz}},
  \bibinfo{author}{\bibfnamefont{P.~M.} \bibnamefont{Oppeneer}},
  \bibinfo{author}{\bibfnamefont{A.~I.} \bibnamefont{Lichtenstein}},
  \bibinfo{author}{\bibfnamefont{M.~I.} \bibnamefont{Katsnelson}},
  \bibnamefont{and} \bibinfo{author}{\bibfnamefont{R.}~\bibnamefont{Caciuffo}},
  \bibinfo{journal}{Phys. Rev. B} \textbf{\bibinfo{volume}{87}},
  \bibinfo{pages}{020505} (\bibinfo{year}{2013}).

\bibitem[{\citenamefont{Joyce et~al.}(2003)\citenamefont{Joyce, Wills,
  Durakiewicz, Butterfield, Guziewicz, Sarrao, Morales, Arko, and
  Eriksson}}]{joyce03}
\bibinfo{author}{\bibfnamefont{J.~J.} \bibnamefont{Joyce}},
  \bibinfo{author}{\bibfnamefont{J.~M.} \bibnamefont{Wills}},
  \bibinfo{author}{\bibfnamefont{T.}~\bibnamefont{Durakiewicz}},
  \bibinfo{author}{\bibfnamefont{M.~T.} \bibnamefont{Butterfield}},
  \bibinfo{author}{\bibfnamefont{E.}~\bibnamefont{Guziewicz}},
  \bibinfo{author}{\bibfnamefont{J.~L.} \bibnamefont{Sarrao}},
  \bibinfo{author}{\bibfnamefont{L.~A.} \bibnamefont{Morales}},
  \bibinfo{author}{\bibfnamefont{A.~J.} \bibnamefont{Arko}}, \bibnamefont{and}
  \bibinfo{author}{\bibfnamefont{O.}~\bibnamefont{Eriksson}},
  \bibinfo{journal}{Phys. Rev. Lett.} \textbf{\bibinfo{volume}{91}},
  \bibinfo{pages}{176401} (\bibinfo{year}{2003}).

\bibitem[{\citenamefont{Eloirdi et~al.}(2009)\citenamefont{Eloirdi, Havela,
  Gouder, Shick, Rebizant, Huber, and Caciuffo}}]{eloirdi09}
\bibinfo{author}{\bibfnamefont{R.}~\bibnamefont{Eloirdi}},
  \bibinfo{author}{\bibfnamefont{L.}~\bibnamefont{Havela}},
  \bibinfo{author}{\bibfnamefont{T.}~\bibnamefont{Gouder}},
  \bibinfo{author}{\bibfnamefont{A.}~\bibnamefont{Shick}},
  \bibinfo{author}{\bibfnamefont{J.}~\bibnamefont{Rebizant}},
  \bibinfo{author}{\bibfnamefont{F.}~\bibnamefont{Huber}}, \bibnamefont{and}
  \bibinfo{author}{\bibfnamefont{R.}~\bibnamefont{Caciuffo}},
  \bibinfo{journal}{J. Nucl. Mater.} \textbf{\bibinfo{volume}{385}},
  \bibinfo{pages}{8 } (\bibinfo{year}{2009}).

\bibitem[{\citenamefont{Ramshaw et~al.}(2015)\citenamefont{Ramshaw, Shekhter,
  McDonald, Betts, Mitchell, Tobash, Mielke, Bauer, and Migliori}}]{ramshaw15}
\bibinfo{author}{\bibfnamefont{B.~J.} \bibnamefont{Ramshaw}},
  \bibinfo{author}{\bibfnamefont{A.}~\bibnamefont{Shekhter}},
  \bibinfo{author}{\bibfnamefont{R.~D.} \bibnamefont{McDonald}},
  \bibinfo{author}{\bibfnamefont{J.~B.} \bibnamefont{Betts}},
  \bibinfo{author}{\bibfnamefont{J.~N.} \bibnamefont{Mitchell}},
  \bibinfo{author}{\bibfnamefont{P.~H.} \bibnamefont{Tobash}},
  \bibinfo{author}{\bibfnamefont{C.~H.} \bibnamefont{Mielke}},
  \bibinfo{author}{\bibfnamefont{E.~D.} \bibnamefont{Bauer}}, \bibnamefont{and}
  \bibinfo{author}{\bibfnamefont{A.}~\bibnamefont{Migliori}},
  \bibinfo{journal}{Proc. Natl. Acad. Sci. U.S.A.} \textbf{\bibinfo{volume}{112}},
  \bibinfo{pages}{3285} (\bibinfo{year}{2015}).

\bibitem[{\citenamefont{Koutroulakis et~al.}(2016)\citenamefont{Koutroulakis,
  Yasuoka, Tobash, Mitchell, Bauer, and Thompson}}]{koutroulakis16}
\bibinfo{author}{\bibfnamefont{G.}~\bibnamefont{Koutroulakis}},
  \bibinfo{author}{\bibfnamefont{H.}~\bibnamefont{Yasuoka}},
  \bibinfo{author}{\bibfnamefont{P.~H.} \bibnamefont{Tobash}},
  \bibinfo{author}{\bibfnamefont{J.~N.} \bibnamefont{Mitchell}},
  \bibinfo{author}{\bibfnamefont{E.~D.} \bibnamefont{Bauer}}, \bibnamefont{and}
  \bibinfo{author}{\bibfnamefont{J.~D.} \bibnamefont{Thompson}},
  \bibinfo{journal}{Phys. Rev. B} \textbf{\bibinfo{volume}{94}},
  \bibinfo{pages}{165115} (\bibinfo{year}{2016}).

\bibitem[{\citenamefont{Eloirdi et~al.}(2017)\citenamefont{Eloirdi, Giacobbe,
  Amador~Celdran, Magnani, Lander, Griveau, Colineau, Miyake, and
  Caciuffo}}]{eloirdi17}
\bibinfo{author}{\bibfnamefont{R.}~\bibnamefont{Eloirdi}},
  \bibinfo{author}{\bibfnamefont{C.}~\bibnamefont{Giacobbe}},
  \bibinfo{author}{\bibfnamefont{P.}~\bibnamefont{Amador~Celdran}},
  \bibinfo{author}{\bibfnamefont{N.}~\bibnamefont{Magnani}},
  \bibinfo{author}{\bibfnamefont{G.~H.} \bibnamefont{Lander}},
  \bibinfo{author}{\bibfnamefont{J.-C.} \bibnamefont{Griveau}},
  \bibinfo{author}{\bibfnamefont{E.}~\bibnamefont{Colineau}},
  \bibinfo{author}{\bibfnamefont{K.}~\bibnamefont{Miyake}}, \bibnamefont{and}
  \bibinfo{author}{\bibfnamefont{R.}~\bibnamefont{Caciuffo}},
  \bibinfo{journal}{Phys. Rev. B} \textbf{\bibinfo{volume}{95}},
  \bibinfo{pages}{094517} (\bibinfo{year}{2017}).

\bibitem[{\citenamefont{Flint et~al.}(2008)\citenamefont{Flint, Shekhter,
  McDonald, Betts, Mitchell, Dzero, and Coleman}}]{flint08}
\bibinfo{author}{\bibfnamefont{R.}~\bibnamefont{Flint}},
  \bibinfo{author}{\bibfnamefont{A.}~\bibnamefont{Shekhter}},
  \bibinfo{author}{\bibfnamefont{R.~D.} \bibnamefont{McDonald}},
  \bibinfo{author}{\bibfnamefont{J.~B.} \bibnamefont{Betts}},
  \bibinfo{author}{\bibfnamefont{J.~N.} \bibnamefont{Mitchell}},
  \bibinfo{author}{\bibfnamefont{M.}~\bibnamefont{Dzero}}, \bibnamefont{and}
  \bibinfo{author}{\bibfnamefont{P.}~\bibnamefont{Coleman}},
  \bibinfo{journal}{Nat. Phys.} \textbf{\bibinfo{volume}{4}},
  \bibinfo{pages}{643} (\bibinfo{year}{2008}).

\bibitem[{\citenamefont{Flint and Coleman}(2010)}]{flint10}
\bibinfo{author}{\bibfnamefont{R.}~\bibnamefont{Flint}} \bibnamefont{and}
  \bibinfo{author}{\bibfnamefont{P.}~\bibnamefont{Coleman}},
  \bibinfo{journal}{Phys. Rev. Lett.} \textbf{\bibinfo{volume}{105}},
  \bibinfo{pages}{246404} (\bibinfo{year}{2010}).

\bibitem[{\citenamefont{Das et~al.}(2015)\citenamefont{Das, Zhu, and
  Graf}}]{graf15}
\bibinfo{author}{\bibfnamefont{T.}~\bibnamefont{Das}},
  \bibinfo{author}{\bibfnamefont{J.-X.} \bibnamefont{Zhu}}, \bibnamefont{and}
  \bibinfo{author}{\bibfnamefont{M.~J.} \bibnamefont{Graf}},
  \bibinfo{journal}{Sci. Rep.} \textbf{\bibinfo{volume}{5}},
  \bibinfo{pages}{8632} (\bibinfo{year}{2015}).

\bibitem[{\citenamefont{Rogalev et~al.}(2001)\citenamefont{Rogalev, Goulon,
  Goulon-Ginet, and Malgrange}}]{rogalev01}
\bibinfo{author}{\bibfnamefont{A.}~\bibnamefont{Rogalev}},
  \bibinfo{author}{\bibfnamefont{J.}~\bibnamefont{Goulon}},
  \bibinfo{author}{\bibfnamefont{C.}~\bibnamefont{Goulon-Ginet}},
  \bibnamefont{and}
  \bibinfo{author}{\bibfnamefont{C.}~\bibnamefont{Malgrange}}, in
  \emph{\bibinfo{booktitle}{Lecture Notes in Physics}}, edited by
  \bibinfo{editor}{\bibfnamefont{E.}~\bibnamefont{Beaurepaire}},
  \bibinfo{editor}{\bibfnamefont{F.}~\bibnamefont{Scheurer}},
  \bibinfo{editor}{\bibfnamefont{G.}~\bibnamefont{Krill}}, \bibnamefont{and}
  \bibinfo{editor}{\bibfnamefont{J.-P.} \bibnamefont{Kappler}}
  (\bibinfo{publisher}{Springer Berlin}, \bibinfo{year}{2001}), vol.
  \bibinfo{volume}{565}, p.~\bibinfo{pages}{60}.

\bibitem[{\citenamefont{Halevy et~al.}(2012)\citenamefont{Halevy, Hen, Orion,
  Colineau, Eloirdi, Griveau, Gaczy\ifmmode~\acute{n}\else \'{n}\fi{}ski,
  Wilhelm, Rogalev, Sanchez et~al.}}]{halevy12}
\bibinfo{author}{\bibfnamefont{I.}~\bibnamefont{Halevy}},
  \bibinfo{author}{\bibfnamefont{A.}~\bibnamefont{Hen}},
  \bibinfo{author}{\bibfnamefont{I.}~\bibnamefont{Orion}},
  \bibinfo{author}{\bibfnamefont{E.}~\bibnamefont{Colineau}},
  \bibinfo{author}{\bibfnamefont{R.}~\bibnamefont{Eloirdi}},
  \bibinfo{author}{\bibfnamefont{J.-C.} \bibnamefont{Griveau}},
  \bibinfo{author}{\bibfnamefont{P.}~\bibnamefont{Gaczy\ifmmode~\acute{n}\else
  \'{n}\fi{}ski}}, \bibinfo{author}{\bibfnamefont{F.}~\bibnamefont{Wilhelm}},
  \bibinfo{author}{\bibfnamefont{A.}~\bibnamefont{Rogalev}},
  \bibinfo{author}{\bibfnamefont{J.-P.} \bibnamefont{Sanchez}},
  \bibinfo{author}{\bibfnamefont{M.~L.}~\bibnamefont{Winterrose}},
  \bibinfo{author}{\bibfnamefont{N.}~\bibnamefont{Magnani}},
  \bibinfo{author}{\bibfnamefont{A.~B.}~\bibnamefont{Shick}}, \bibnamefont{and}
  \bibinfo{author}{\bibfnamefont{R.}~\bibnamefont{Caciuffo}},
  \bibinfo{journal}{Phys. Rev. B}
  \textbf{\bibinfo{volume}{85}}, \bibinfo{pages}{014434}
  (\bibinfo{year}{2012}).

\bibitem[{\citenamefont{Magnani et~al.}(2015)\citenamefont{Magnani, Caciuffo,
  Wilhelm, Colineau, Eloirdi, Griveau, Rusz, Oppeneer, Rogalev, and
  Lander}}]{magnani15}
\bibinfo{author}{\bibfnamefont{N.}~\bibnamefont{Magnani}},
  \bibinfo{author}{\bibfnamefont{R.}~\bibnamefont{Caciuffo}},
  \bibinfo{author}{\bibfnamefont{F.}~\bibnamefont{Wilhelm}},
  \bibinfo{author}{\bibfnamefont{E.}~\bibnamefont{Colineau}},
  \bibinfo{author}{\bibfnamefont{R.}~\bibnamefont{Eloirdi}},
  \bibinfo{author}{\bibfnamefont{J.-C.} \bibnamefont{Griveau}},
  \bibinfo{author}{\bibfnamefont{J.}~\bibnamefont{Rusz}},
  \bibinfo{author}{\bibfnamefont{P.~M.} \bibnamefont{Oppeneer}},
  \bibinfo{author}{\bibfnamefont{A.}~\bibnamefont{Rogalev}}, \bibnamefont{and}
  \bibinfo{author}{\bibfnamefont{G.~H.} \bibnamefont{Lander}},
  \bibinfo{journal}{Phys. Rev. Lett.} \textbf{\bibinfo{volume}{114}},
  \bibinfo{pages}{097203} (\bibinfo{year}{2015}).

\bibitem[{\citenamefont{Khomskii}(1979)}]{khomskii79}
\bibinfo{author}{\bibfnamefont{D.}~\bibnamefont{Khomskii}},
  \bibinfo{journal}{Sov. Phys. Usp.} \textbf{\bibinfo{volume}{22}},
  \bibinfo{pages}{879} (\bibinfo{year}{1979}).

\bibitem[{\citenamefont{Wilhelm et~al.}(2013)\citenamefont{Wilhelm, Eloirdi,
  Rusz, Springell, Colineau, Griveau, Oppeneer, Caciuffo, Rogalev, and
  Lander}}]{wilhelm13}
\bibinfo{author}{\bibfnamefont{F.}~\bibnamefont{Wilhelm}},
  \bibinfo{author}{\bibfnamefont{R.}~\bibnamefont{Eloirdi}},
  \bibinfo{author}{\bibfnamefont{J.}~\bibnamefont{Rusz}},
  \bibinfo{author}{\bibfnamefont{R.}~\bibnamefont{Springell}},
  \bibinfo{author}{\bibfnamefont{E.}~\bibnamefont{Colineau}},
  \bibinfo{author}{\bibfnamefont{J.-C.} \bibnamefont{Griveau}},
  \bibinfo{author}{\bibfnamefont{P.~M.} \bibnamefont{Oppeneer}},
  \bibinfo{author}{\bibfnamefont{R.}~\bibnamefont{Caciuffo}},
  \bibinfo{author}{\bibfnamefont{A.}~\bibnamefont{Rogalev}}, \bibnamefont{and}
  \bibinfo{author}{\bibfnamefont{G.~H.} \bibnamefont{Lander}},
  \bibinfo{journal}{Phys. Rev. B} \textbf{\bibinfo{volume}{88}},
  \bibinfo{pages}{024424} (\bibinfo{year}{2013}).

\bibitem[{\citenamefont{Ohishi et~al.}(2007)\citenamefont{Ohishi, Heffner,
  Morris, Bauer, Graf, Zhu, Morales, Sarrao, Fluss, MacLaughlin
  et~al.}}]{ohishi07}
\bibinfo{author}{\bibfnamefont{K.}~\bibnamefont{Ohishi}},
  \bibinfo{author}{\bibfnamefont{R.~H.} \bibnamefont{Heffner}},
  \bibinfo{author}{\bibfnamefont{G.~D.} \bibnamefont{Morris}},
  \bibinfo{author}{\bibfnamefont{E.~D.} \bibnamefont{Bauer}},
  \bibinfo{author}{\bibfnamefont{M.~J.} \bibnamefont{Graf}},
  \bibinfo{author}{\bibfnamefont{J.-X.} \bibnamefont{Zhu}},
  \bibinfo{author}{\bibfnamefont{L.~A.} \bibnamefont{Morales}},
  \bibinfo{author}{\bibfnamefont{J.~L.} \bibnamefont{Sarrao}},
  \bibinfo{author}{\bibfnamefont{M.~J.} \bibnamefont{Fluss}},
  \bibinfo{author}{\bibfnamefont{D.~E.} \bibnamefont{MacLaughlin}},
  \bibinfo{author}{\bibfnamefont{L.} \bibnamefont{Shu}},
  \bibinfo{author}{\bibfnamefont{W.} \bibnamefont{Higemoto}}, \bibnamefont{and}
  \bibinfo{author}{\bibfnamefont{T.~U.} \bibnamefont{Ito}},
  \bibinfo{journal}{Phys. Rev. B}
  \textbf{\bibinfo{volume}{76}}, \bibinfo{pages}{064504}
  (\bibinfo{year}{2007}).

\bibitem[{\citenamefont{Sonier et~al.}(1997)\citenamefont{Sonier, Kiefl,
  Brewer, Bonn, Dunsiger, Hardy, Liang, MacFarlane, Riseman, Noakes
  et~al.}}]{sonier97}
\bibinfo{author}{\bibfnamefont{J.~E.} \bibnamefont{Sonier}},
  \bibinfo{author}{\bibfnamefont{R.~F.} \bibnamefont{Kiefl}},
  \bibinfo{author}{\bibfnamefont{J.~H.} \bibnamefont{Brewer}},
  \bibinfo{author}{\bibfnamefont{D.~A.} \bibnamefont{Bonn}},
  \bibinfo{author}{\bibfnamefont{S.~R.} \bibnamefont{Dunsiger}},
  \bibinfo{author}{\bibfnamefont{W.~N.} \bibnamefont{Hardy}},
  \bibinfo{author}{\bibfnamefont{R.}~\bibnamefont{Liang}},
  \bibinfo{author}{\bibfnamefont{W.~A.} \bibnamefont{MacFarlane}},
  \bibinfo{author}{\bibfnamefont{T.~M.} \bibnamefont{Riseman}},
  \bibinfo{author}{\bibfnamefont{D.~R.} \bibnamefont{Noakes}}, \bibnamefont{and}
  \bibinfo{author}{\bibfnamefont{C.~E.} \bibnamefont{Stronach}},
  \bibinfo{journal}{Phys. Rev. B}
  \textbf{\bibinfo{volume}{55}}, \bibinfo{pages}{11789} (\bibinfo{year}{1997}).

\bibitem[{\citenamefont{Ummarino
  et~al.}(2009{\natexlab{b}})\citenamefont{Ummarino, Magnani, Colineau,
  Griveau, Jutier, Rebizant, Wastin, and Caciuffo}}]{ummarino09b}
\bibinfo{author}{\bibfnamefont{G.}~\bibnamefont{Ummarino}},
  \bibinfo{author}{\bibfnamefont{N.}~\bibnamefont{Magnani}},
  \bibinfo{author}{\bibfnamefont{E.}~\bibnamefont{Colineau}},
  \bibinfo{author}{\bibfnamefont{J.-C.} \bibnamefont{Griveau}},
  \bibinfo{author}{\bibfnamefont{F.}~\bibnamefont{Jutier}},
  \bibinfo{author}{\bibfnamefont{J.}~\bibnamefont{Rebizant}},
  \bibinfo{author}{\bibfnamefont{F.}~\bibnamefont{Wastin}}, \bibnamefont{and}
  \bibinfo{author}{\bibfnamefont{R.}~\bibnamefont{Caciuffo}},
  \bibinfo{journal}{Physica B: Condens. Matter} \textbf{\bibinfo{volume}{404}},
  \bibinfo{pages}{3213 } (\bibinfo{year}{2009}{\natexlab{b}}).

\bibitem[{\citenamefont{Thole and van~der Laan}(1988)}]{thole88}
\bibinfo{author}{\bibfnamefont{B.~T.} \bibnamefont{Thole}} \bibnamefont{and}
  \bibinfo{author}{\bibfnamefont{G.}~\bibnamefont{van~der Laan}},
  \bibinfo{journal}{Phys. Rev. A} \textbf{\bibinfo{volume}{38}},
  \bibinfo{pages}{1943} (\bibinfo{year}{1988}).

\bibitem[{\citenamefont{van~der Laan et~al.}(2004)\citenamefont{van~der Laan,
  Moore, Tobin, Chung, Wall, and Schwartz}}]{vanderlaan04}
\bibinfo{author}{\bibfnamefont{G.}~\bibnamefont{van~der Laan}},
  \bibinfo{author}{\bibfnamefont{K.~T.} \bibnamefont{Moore}},
  \bibinfo{author}{\bibfnamefont{J.~G.} \bibnamefont{Tobin}},
  \bibinfo{author}{\bibfnamefont{B.~W.} \bibnamefont{Chung}},
  \bibinfo{author}{\bibfnamefont{M.~A.} \bibnamefont{Wall}}, \bibnamefont{and}
  \bibinfo{author}{\bibfnamefont{A.~J.} \bibnamefont{Schwartz}},
  \bibinfo{journal}{Phys. Rev. Lett.} \textbf{\bibinfo{volume}{93}},
  \bibinfo{pages}{097401} (\bibinfo{year}{2004}).

\bibitem[{\citenamefont{Janoschek et~al.}(2015)\citenamefont{Janoschek, Haskel,
  Fernandez-Rodriguez, van Veenendaal, Rebizant, Lander, Zhu, Thompson, and
  Bauer}}]{janoschek15}
\bibinfo{author}{\bibfnamefont{M.}~\bibnamefont{Janoschek}},
  \bibinfo{author}{\bibfnamefont{D.}~\bibnamefont{Haskel}},
  \bibinfo{author}{\bibfnamefont{J.}~\bibnamefont{Fernandez-Rodriguez}},
  \bibinfo{author}{\bibfnamefont{M.}~\bibnamefont{van Veenendaal}},
  \bibinfo{author}{\bibfnamefont{J.}~\bibnamefont{Rebizant}},
  \bibinfo{author}{\bibfnamefont{G.~H.} \bibnamefont{Lander}},
  \bibinfo{author}{\bibfnamefont{J.-X.} \bibnamefont{Zhu}},
  \bibinfo{author}{\bibfnamefont{J.~D.} \bibnamefont{Thompson}},
  \bibnamefont{and} \bibinfo{author}{\bibfnamefont{E.~D.} \bibnamefont{Bauer}},
  \bibinfo{journal}{Phys. Rev. B} \textbf{\bibinfo{volume}{91}},
  \bibinfo{pages}{035117} (\bibinfo{year}{2015}).

\bibitem[{\citenamefont{Thole et~al.}(1992)\citenamefont{Thole, Carra, Sette,
  and van~der Laan}}]{thole92}
\bibinfo{author}{\bibfnamefont{B.~T.} \bibnamefont{Thole}},
  \bibinfo{author}{\bibfnamefont{P.}~\bibnamefont{Carra}},
  \bibinfo{author}{\bibfnamefont{F.}~\bibnamefont{Sette}}, \bibnamefont{and}
  \bibinfo{author}{\bibfnamefont{G.}~\bibnamefont{van~der Laan}},
  \bibinfo{journal}{Phys. Rev. Lett.} \textbf{\bibinfo{volume}{68}},
  \bibinfo{pages}{1943} (\bibinfo{year}{1992}).

\bibitem[{\citenamefont{Carra et~al.}(1993)\citenamefont{Carra, Thole,
  Altarelli, and Wang}}]{carra93}
\bibinfo{author}{\bibfnamefont{P.}~\bibnamefont{Carra}},
  \bibinfo{author}{\bibfnamefont{B.~T.} \bibnamefont{Thole}},
  \bibinfo{author}{\bibfnamefont{M.}~\bibnamefont{Altarelli}},
  \bibnamefont{and} \bibinfo{author}{\bibfnamefont{X.}~\bibnamefont{Wang}},
  \bibinfo{journal}{Phys. Rev. Lett.} \textbf{\bibinfo{volume}{70}},
  \bibinfo{pages}{694} (\bibinfo{year}{1993}).

\bibitem[{\citenamefont{Kappler et~al.}(2004)\citenamefont{Kappler, Herr,
  Schmerber, Derory, Parlebas, Jaouen, Wilhelm, and Rogalev}}]{kappler04}
\bibinfo{author}{\bibfnamefont{J.-P.} \bibnamefont{Kappler}},
  \bibinfo{author}{\bibfnamefont{A.}~\bibnamefont{Herr}},
  \bibinfo{author}{\bibfnamefont{G.}~\bibnamefont{Schmerber}},
  \bibinfo{author}{\bibfnamefont{A.}~\bibnamefont{Derory}},
  \bibinfo{author}{\bibfnamefont{J.-C.} \bibnamefont{Parlebas}},
  \bibinfo{author}{\bibfnamefont{N.}~\bibnamefont{Jaouen}},
  \bibinfo{author}{\bibfnamefont{F.}~\bibnamefont{Wilhelm}}, \bibnamefont{and}
  \bibinfo{author}{\bibfnamefont{A.}~\bibnamefont{Rogalev}},
  \bibinfo{journal}{Eur. Phys. J. B} \textbf{\bibinfo{volume}{37}},
  \bibinfo{pages}{163} (\bibinfo{year}{2004}).

\bibitem[{\citenamefont{Opahle and Oppeneer}(2003)}]{opahle03}
\bibinfo{author}{\bibfnamefont{I.}~\bibnamefont{Opahle}} \bibnamefont{and}
  \bibinfo{author}{\bibfnamefont{P.~M.} \bibnamefont{Oppeneer}},
  \bibinfo{journal}{Phys. Rev. Lett.} \textbf{\bibinfo{volume}{90}},
  \bibinfo{pages}{157001} (\bibinfo{year}{2003}).

\bibitem[{\citenamefont{Luca et~al.}(2010)\citenamefont{Luca, Ghiringhelli,
  Sala, Matteo, Haverkort, Berger, Bisogni, Cezar, Brookes, and
  Salluzzo}}]{deluca10}
\bibinfo{author}{\bibfnamefont{G.~M.} \bibnamefont{De~Luca}},
  \bibinfo{author}{\bibfnamefont{G.}~\bibnamefont{Ghiringhelli}},
  \bibinfo{author}{\bibfnamefont{M.} \bibnamefont{Moretti~Sala}},
  \bibinfo{author}{\bibfnamefont{S.} \bibnamefont{Di~Matteo}},
  \bibinfo{author}{\bibfnamefont{M.~W.} \bibnamefont{Haverkort}},
  \bibinfo{author}{\bibfnamefont{H.}~\bibnamefont{Berger}},
  \bibinfo{author}{\bibfnamefont{V.}~\bibnamefont{Bisogni}},
  \bibinfo{author}{\bibfnamefont{J.~C.} \bibnamefont{Cezar}},
  \bibinfo{author}{\bibfnamefont{N.~B.} \bibnamefont{Brookes}},
  \bibnamefont{and} \bibinfo{author}{\bibfnamefont{M.}~\bibnamefont{Salluzzo}},
  \bibinfo{journal}{Phys. Rev. B} \textbf{\bibinfo{volume}{82}},
  \bibinfo{pages}{214504} (\bibinfo{year}{2010}).

\end{thebibliography}

\end{document}